\documentclass[aoas,MSNbibl,nameyear,dvips]{arximspdf}
\usepackage{graphicx}

\doi{10.1214/12-AOAS554} 
\volume{6}
\issue{4}
\pubyear{2012}
\firstpage{1406}
\lastpage{1429}

\makeatletter

\newtheorem{proposition}{Proposition}

\newproclaim{remark}{Remark}
\newproclaim{example}{Example}

\newcommand{\cal}{\mathcal}
\newcommand{\mathbbm}{\mathbf}

\newcommand{\reels}{\mathbb R}
\newcommand{\Proba}{\mathbb P}

\makeatother

\begin{document}
\begin{frontmatter}

\title{Approximating the conditional density given large observed
values via a multivariate extremes framework, with application
to~environmental~data}
\runtitle{Conditional density approximation for extremes}

\begin{aug}
\author[A]{\fnms{Daniel} \snm{Cooley}\corref{}\thanksref{t1}\ead[label=e1]{cooleyd@stat.colostate.edu}},
\author[B]{\fnms{Richard A.} \snm{Davis}\thanksref{t2}}
\and
\author[C]{\fnms{Philippe} \snm{Naveau}\thanksref{t3}}
\runauthor{D. Cooley, R. A. Davis and P. Naveau}
\affiliation{Colorado State University, Columbia University and LSCE-CNRS}
\address[A]{D. Cooley\\
Department of Statistics\\
Colorado State University\\
Fort Collins, Colorado\\
USA\\
\printead{e1}} 
\address[B]{R. A. Davis\\
Department of Statistics\\
Columbia University\\
New York, New York\\
USA}
\address[C]{P. Naveau\\
Laboratoire des Sciences du Climat\\
\quad et de l'Environnement\\
IPSL-CNRS\\
Gif-sur-Yvette\\
France}
\end{aug}

\thankstext{t1}{Supported in part by NSF Grant DMS-09-05315.}

\thankstext{t2}{Supported in part by NSF Grants DMS-07-43459 and
DMS-11-07031.}

\thankstext{t3}{Supported in part by the EU\_FP7 ACQWA project
(\protect\url{http://www.acqwa.ch}) under contract 212250,
by the ANR-MOPERA, ANR-Mc Sim, MIRACCLE-GICC, PEPER-GIS projects.}

\received{\smonth{5} \syear{2011}}
\revised{\smonth{3} \syear{2012}}

%
\begin{abstract}
%
Phenomena such as air pollution levels are of greatest interest when
observations are large, but standard prediction methods are not
specifically designed for large observations.
We propose a method, rooted in extreme value theory, which approximates
the conditional distribution of an unobserved component of a random
vector given large observed values.
Specifically, for $\mathbbm Z = (Z_1, \ldots, Z_d)^T$ and $\mathbbm Z_{-d} =
(Z_1, \ldots, Z_{d-1})^T$, the method approximates the conditional
distribution of $[Z_d | \mathbbm Z_{-d} = \mathbbm z_{-d}]$ when $ \| \mathbbm z_{-d}
\| > r_*$.
The approach is based on the assumption that $\mathbbm Z$ is a multivariate
regularly varying random vector of dimension $d$.
The conditional distribution approximation relies on knowledge of the
angular measure of $\mathbbm Z$, which provides explicit structure for
dependence in the distribution's tail.
As the method produces a predictive distribution rather than just a
point predictor, one can answer any question posed about the quantity
being predicted, and, in particular, one can assess how well the
extreme behavior is represented.

Using a fitted model for the angular measure, we apply our method to
nitrogen dioxide measurements in metropolitan Washington DC.
We obtain a predictive distribution for the air pollutant at a location
given the air pollutant's measurements at four nearby locations and
given that the norm of the vector of the observed measurements is large.
\end{abstract}

%

\end{frontmatter}

\section{Introduction and motivation}

\begin{figure}

\includegraphics{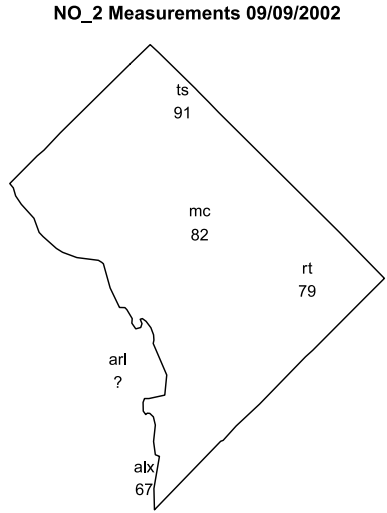}

\caption{Locations of the NO$_2$ monitors used in the Washington DC
study. Locations are Alexandria (alx), McMillan (mc), River Terrace
(rt), Takoma School (ts) and Arlington (arl). Also shown is the
boundary of the District of Columbia. Measurements for September 9,
2002 are shown for the four locations we use for predicting the
measurement at Arlington.}
\label{figmonitorLocs}
\end{figure}

Nitrogen dioxide (NO$_2$) is an air pollutant which is among those
monitored by the US Environmental Protection Agency (EPA).
Figure~\ref{figmonitorLocs} shows NO$_2$ measurements at four
locations in the Washington DC metropolitan area on September 9, 2002.
This day's measurements are particularly large: each of the four
measurements exceeds the 0.97 quantile of the empirical distribution
for its location.
Certainly, air pollution levels are of most interest when pollution
levels are high.
It is natural to ask, given the measurements at these four locations
and given that they are large, what can be said about pollution levels
at nearby unmonitored locations?

Linear prediction methods are questionable when the data are
non-Gaus\-sian, and a better approach may be to approximate the
conditional density.
Extreme value theory leads one to describe the joint tail with
non-Gaussian distributions, and dependence in the tail is typically not
well described by covariances upon which linear prediction methods rely.
In applications like the above air pollution example where interest
lies in the large occurrences, approximating a conditional density
allows one to answer important questions such as ``Given that nearby
locations' measurements are large, what is the probability a certain
unmonitored location exceeds some critical level?'' or ``Given that
nearby locations' measurements are large, what is a reasonable
probabilistic upper bound for the air pollution level at an unmonitored
location?''
Our method to approximate the conditional density is based on extreme
value theory and is therefore specifically designed for instances when
the observations are large.
Extreme value theory provides a framework for describing the dependence
found in the joint upper tail of the distribution, and, at the same
time, does not require knowledge of the entire joint distribution.
In particular, we will assume that the joint distribution of the
observations and the random variable we wish to predict are
multivariate regularly varying, and we use the angular measure of this
random vector to approximate the conditional density.
By approximating the conditional density, we are able to address
questions of the type posed above about the unobserved random variable.

In the next section we provide some necessary background on extreme
value theory.
In Section~\ref{secpredForExtremes} we discuss prediction for
extremes; we review previous related work in Section~\ref{secrefSection} and then introduce our method in Section~\ref{secpredViaAngMsr}.
In Section~\ref{secapp} we apply the prediction method to the
Washington DC air pollution data.
We conclude with a summary and discussion section.

\section{Characterizing extremes, multivariate regular variation and
the angular measure}
\label{secextBackground}

Extreme value analysis is the branch of statistics and probability
theory whose aim is to describe the upper tail of a distribution.
In this section we give a very brief overview of the discipline,
particularly focusing on multivariate regular variation and the angular measure.
There are a number of excellent resources if one wishes to delve more
into the theory or practice of extreme value analysis.
The book by \citet{deHaan06} gives a comprehensive overview of extreme
value theory in the univariate, multivariate and stochastic process settings.
\citet{Beirlant04} also give a thorough treatment of the theory and give
a broad overview of recent statistical practice.
\citet{Resnick07} focuses on the heavy-tailed case for both the
univariate and multivariate settings.
\citet{Coles01a} gives an approachable introduction to statistical
practice focusing primarily on maximum likelihood inference.

\subsection{Extreme value analysis}

Extreme value analysis is founded on asymptotic results that
characterize a distribution's upper tail by a limited class of functions.
Statistical practice fits this class of functions to a subset of data
which are considered extreme.
Two approaches for choosing this subset of extreme data are commonly
used: in the first, block (e.g., annual) maxima are extracted, in the
second, observations which exceed a threshold are retained.

In the univariate case, asymptotic results lead one to model block
maximum data with a generalized extreme value (GEV) distribution which
consolidates the three extremal types [\citet{Fisher28}, \citet{Gnedenko43}]
into one parametric family.
Threshold exceedance data are generally modeled via the generalized
Pareto distribution (GPD) or an equivalent point process representation.

Statistical modeling of multivariate extremes is more complicated.
Given a sequence of i.i.d. random vectors $\mathbbm Y_i = (Y_{i,1}, \ldots,
Y_{i,d})^T, i = 1, 2, \ldots\,$,
classical multivariate extreme value theory considers the vector of
renormalized element-wise maxima $\frac{\mathbbm M_n - \mathbbm c_n}{\mathbbm b_n}$,
where the division is taken to be element-wise, $\mathbbm M_n = (\bigvee_{i
= 1}^n Y_{i,1}, \ldots, \bigvee_{i = 1}^n Y_{i,d} )^T$, and $\bigvee$
denotes the maximum function.
The theory shows the distribution of $\frac{\mathbbm M_n - \mathbbm c_n}{\mathbbm
b_n}$ converges to a multivariate max-stable distribution
(equivalently, multivariate extreme value distribution), which we
characterize below.
For threshold exceedance data, one must first define what it means for
a random vector to exceed a threshold.
\citet{Rootzen06} define a multivariate GPD which is well-suited to
describe threshold exceedances in which the threshold has been defined
for each vector element.
For the work below, we employ the framework of multivariate regular
variation [\citet{Resnick07}] to describe threshold exceedances.

\subsection{Multivariate regular variation}
\label{secmvRegVar}

Multivariate regular variation is a notion that is used for modeling
multivariate heavy-tailed data.
This behavior is best seen via a natural decomposition into
pseudo-polar coordinates.
If nonnegative $\mathbbm Z = (Z_1, \ldots, Z_d)^T$ is a multivariate
regularly varying random vector, then the radial component $\| \mathbbm Z \|
$ decays like a power function; that is,
$\Proba( \| \mathbbm Z \| > t) = L(t) t^{-\alpha}$, where $L(t)$ is a slowly
varying function\setcounter{footnote}{3}\footnote{$L(t)$ is slowly
varying if $\lim_{t
\rightarrow\infty} \frac{L(st)}{L(t)} = 1$. Roughly, $L(t)$ cannot go
to zero or infinity faster than any power function.} at $\infty$ and
$\alpha> 0$ is termed the tail index.
The angular component, $\| \mathbbm Z \|^{-1} \mathbbm Z$, is described by a
probability measure that lives on the unit sphere and which becomes
independent of the radial component as the radial component drifts off
to infinity.
Central ideas in the more detailed treatment that follows are: (1) the
convergence to a measure $\Lambda$ that serves as the intensity measure
for a limiting point process, (2) that the limiting intensity measure
is a product measure when described in terms of radial and angular
components, and (3) the angular measure $H$ which describes the
distribution of the angular components.

There are several equivalent definitions of multivariate regular
variation of a random vector.
We say the nonnegative random vector $\mathbbm Z$ is regularly varying if
%
\begin{equation}
\label{eqnmvRegVar1}
\frac
{\Proba(t^{-1} \mathbbm Z \in\cdot)}
{\Proba(\| \mathbbm Z \| > t)}
\stackrel{v}{\longrightarrow}
\Lambda(\cdot)
\end{equation}
as $t \rightarrow\infty$, where $v$ denotes vague convergence on
${\cal C} = [0,\infty]^d \setminus\{\mathbbm0\}$ and $\| \cdot\|$ is any
norm\footnote{The compact sets in ${\cal C}$ are all closed sets in
$[0,\infty]^d$, that do not contain $\mathbbm0$, that is, the closed sets
bounded away from $\mathbbm0$.}
[\citet{Resnick07}, Chapter 3].
For any measureable set $A \subset{\cal C}$ and scalar $s > 0$, the
measure $\Lambda$ has the scaling property
%
\begin{equation}
\label{eqnscaling}
\Lambda(sA) = s^{-\alpha} \Lambda(A)
\end{equation}
from which one sees the power-function-like behavior.
Choosing a sequence $a_n$ such $\Proba(\| \mathbbm Z \| > a_n) \sim n^{-1}$,
one can obtain the sequential version of
(\ref{eqnmvRegVar1})\footnote{Other normalizing
sequences are sometimes used; see \citet{Resnick07},
page 174.}:
%
\begin{equation}
\label{eqnmvRegVarSeq}
n \Proba \biggl( \frac{\mathbbm Z}{a_n} \in\cdot\biggr) \stackrel{v}{\longrightarrow
} \Lambda(\cdot).
\end{equation}
%

The transformation to polar coordinates $R = \| \mathbbm Z \|$ and $\mathbbm W =
\| \mathbbm Z \|^{-1} \mathbbm Z$ naturally arises from the scaling property (\ref
{eqnscaling}).
If $S_{d-1} = \{ \mathbbm z \in{\cal C} \mid\| \mathbbm z \| = 1 \}$ denotes
the unit sphere under the chosen norm,
then one can show there exists a probability measure $H$ on $S_{d-1}$
such that for any $H$-continuity Borel subset $B$ of $S_{d-1}$,
%
\begin{equation}
\label{eqnmvRegVarPolar}
n \Proba\biggl( \frac{R}{a_n} > r, \mathbbm W \in B \biggr) 
{\longrightarrow} r^{-\alpha} H(B).
\end{equation}
Following \citet{Resnick07}, we refer to $H$ as the angular measure,
although it is also referred to as the spectral measure.
The advantage of the polar transformation is that the radial component
$R$ acts independently of the angular component $\mathbbm W$ whose behavior
is captured by $H$.

From (\ref{eqnmvRegVarPolar}), one can characterize the tail behavior
of $\mathbbm Z$ if one knows (or can estimate) $\alpha$ and $H$.
However, without further assumptions, this proves to be difficult, as
$H$ can be any probability measure on $S_{d-1}$.
For simplification purposes, in multivariate extremes it is often
assumed that the components $Z_j, j = 1, \ldots, d$, of the random
vector have a common marginal distribution, not just the common tail
index that is implied by the general conditions of multivariate regular
variation [e.g., \citet{resnick2002}, Section 2].
There is no loss in generality by assuming specific margins [\citet
{resnick87}, Proposition 5.10].
For the remainder, we will assume $\mathbbm Z = (Z_1, \ldots, Z_d)^T$ is
regularly varying with tail index $\alpha= 1$ and that $Z_j, j = 1,
\ldots, d$, have a common marginal distribution.
Under this assumption, it follows that
%
\begin{equation}
\label{eqncomCondition}
\int_{S_{d-1}} w_1 \,dH(\mathbbm w) = \int_{S_{d-1}} w_j \,dH(\mathbbm w)
\qquad\mbox{for } j = 2, \ldots, d,
\end{equation}
providing some structure to the angular measure $H$.
Furthermore, when $\alpha= 1$ it is particularly useful to choose the
$L_1$ norm: $\| \mathbbm z \| = z_1 + \cdots+ z_d$, for which the unit
sphere is the simplex $S_{d-1} = \{\mathbbm w \in{\cal C}\dvtx w_1 + \cdots+
w_d = 1\}$.
With this norm, $1 = \int_{S_{d-1}} \,dH(\mathbbm w) = \int_{S_{d-1}} (w_1 +
\cdots+ w_d) \,dH(\mathbbm w)$ and, hence, $\int_{S_{d-1}} w_j \,dH(\mathbbm w) = d^{-1}$.
In practice, the assumption of common marginals (or, for that matter, a
common tail index) is rarely met.
If the data that one intends to model arise from a $d$-dimensional
random vector $\mathbbm Y$ for which the regularly varying $\alpha= 1$ and
common marginal assumptions do not hold, we presume there exist
probability integral transforms $T_j$ such that $T_j(Y_j) = Z_j$ for $j
= 1, \ldots, d$.
This preprocessing of the random variables is common in extreme value
analyses [e.g., \citet{cooley10b}, \citet{coles91}] and can be viewed
analogously to the preprocessing required to fit a stationary model to
time series or spatial data.

One recognizes that (\ref{eqnmvRegVarSeq}) 
is the classic relationship characterizing convergence to a Poisson
process, and it is often useful to think in terms of point processes
when describing multivariate regular variation. From
(\ref{eqnmvRegVarSeq}), the sequence of point processes $N_n$
consisting of point masses located at $\mathbbm Z_i/a_n, i = 1, 2,
\ldots, n$, where $\mathbbm Z_i$ are i.i.d. copies of $\mathbbm Z$
converges to a nonhomogeneous Poisson process $N$ with intensity
measure $\Lambda(\cdot)$ on ${\mathbb B}({\cal C})$
[\citet{Resnick07}, Section 6.2].
We denote the corresponding intensity function by $\Lambda(d \mathbbm
z)$, where $\Lambda(A) = \int_A \Lambda(d \mathbbm z)$.
From~(\ref{eqnmvRegVarPolar}), in terms of polar coordinates, $\Lambda
(dr \times d\mathbbm w) = r^{-2} \,dr H( d\mathbbm w)$. If the angular
measure $H$ is differentiable, then we refer to the angular density
$h(\mathbbm w)$, and $\Lambda(dr \times d\mathbbm w) = r^{-2}
h(\mathbbm w) \,dr \,d\mathbbm w$.

Multivariate regular variation is a useful way to characterize the
joint upper tail of a random variable $\mathbbm Y$ for a number of reasons.
First, interest in extreme behavior is often greatest in cases when the
tails are believed to be heavy (i.e., having asymptotic behavior like a
power function), and multivariate regular variation provides a
mathematical framework for such behavior.
Even when the tails do not share a common tail index, marginal
transformations as described above can be employed to utilize the framework.
Second and more importantly, the angular measure $H$ specifically
describes the dependence found in the tails.
Since our interest is in performing prediction when the observations
are large, it is natural to use a framework specifically designed for
describing tail dependence.

We perform prediction by approximating the conditional density.
To do so, we will rely on a model for the angular measure $H$, and this
model must be able to be evaluated for any $\mathbbm w \in S_{d-1}$.
There have been several parametric models proposed for $H$ which meet
the moment conditions (\ref{eqncomCondition}).
An early parametric model was the tilted Dirichlet model of \citet{coles91}.
Recently, \citet{cooley10b} and \citet{Ballani11} employed a geometric
approach to construct new parametric models.
A semi-parametric model via a mixture of Dirichlet densities was
introduced by \citet{Boldi07}.
Model fitting is done by \citet{coles91}, \citet{cooley10b} and \citet
{Ballani11} via a likelihood based on
the point process representation for multivariate regular variation,
while \citet{Boldi07} use both a Bayesian MCMC approach as well as an EM
approach to fit their mixture model.
Once fit, any of these models could be used for $H$ in the prediction
procedure we outline in Section
\ref{secpredViaAngMsr}.


There is further justification for using the framework of multivariate
regular variation for modeling extreme values.
The multivariate max-stable distributions obtained by the classical
theory can be characterized by the angular measure $H$.
If one assumes that the marginals of the limiting distribution are unit
Fr\'echet [$\Proba(Z \leq z) = \exp(-z^{-1})$]; that is, the domain
of attraction of all regularly varying random variables with $\alpha=
1$, then
%
\begin{equation}
\label{eqnmvevdAngMsr}
\Proba\biggl(\frac{\mathbbm M_n}{b_n} \leq\mathbbm z\biggr)
\stackrel{d}{\longrightarrow}
\exp\Biggl[ -d \int_{S_{d-1}} \bigvee_{j = 1}^d
\biggl(\frac{w_j}{z_j} \biggr) \,dH(\mathbbm w) \Biggr].
\end{equation}
Here the normalizing sequence $b_n = a_n/d$ to obtain the unit-Fr\'
echet marginals.
There have been parametric models developed which give closed-form
expressions for subfamilies of multivariate max-stable distributions
such as the asymmetric logistic [\citet{Tawn90}] and negative logistic
[\citet{Joe90}], and these can be used to fit block maxima.
Besides being max-stable, these distributions are multivariate
regularly varying and we later use the logistic model [\citet{Gumbel60}]
to simulate random vectors whose distribution function and limiting
angular measure are both known in closed form.


%

\section{Conditional distribution estimation and prediction for extremes}
\label{secpredForExtremes}

\subsection{Previous work in prediction for extremes}
\label{secrefSection}

There has been a small amount of work which has tried to devise methods
for performing prediction for extremes.
Davis and Resnick (\citeyear{Davis89,Davis93}) define a distance $d$ between the components of
a bivariate max-stable random variable, and suggest a method of
prediction which minimizes the distance between the observed component
and the predictor.
\citet{Craigmile06} offer a geostatistical approach to the problem of
determining exceedances in a spatial setting by adjusting the loss
function of the kriging predictor.

A recent important advance in the area of approximating a conditional
distribution for extremes is the work of \citet{wang2011}, and we view
the work in this paper as complementary. 
Wang and Stoev perform prediction for the case of max-stable random vectors.
Let $\mathbbm M_n^{(d+p)} = (\mathbbm M_n^{(d)}, \mathbbm M_n^{(p)})^T$, where $\mathbbm
M_n^{(d)} = (M_{n,1}, \ldots, M_{n, d})^T$, $\mathbbm M_n^{(p)} = (M_{n,1},
\ldots, M_{n, p})^T$, and where\break
$\mathbbm M_n^{(d+p)}$ is assumed to be a max-stable random vector with a
known distribution.
Given data $\mathbbm m_n^{(d)} = (m_{n,1}, \ldots, m_{n, d})^T$, Wang and
Stoev obtain approximate draws from $\mathbbm M_n^{(p)} \mid\mathbbm M_n^{(d)} =
\mathbbm m_n^{(d)}$.
They accomplish this by sampling from spectrally-discrete max-stable
models which can be represented as a max-linear combination of
independent random variables.
Using a spectrally discrete model would seem to be limiting, as it
would imply that the corresponding angular measure would only have mass
at discrete locations.
However, it is known that any multivariate max-stable distribution can
be approximated arbitrarily well by a max-linear model with a
sufficient number of elements, and Wang and Stoev claim that their
computational method can handle max-linear combinations on the order of
thousands.
\citet{wang2011} apply their approach in the spatial setting and the
results show the discrete approximation performs quite well.

The method we propose in the next section differs from that of \citet
{wang2011} in a number of important ways.
Perhaps the most important difference is that, rather than performing
prediction in a max-stable setting which would lend itself to data that
are block maxima, our prediction method is best suited for large
observations, that is, the threshold exceedance case.
Another difference is that, rather than successively drawing from the
conditional distribution as Wang and Stoev do, we provide an analytic
approximation to the conditional density given the observations are
sufficiently large.
Additionally, rather than relying on an approximation which corresponds
to a discrete angular measure, our method instead relies on a
parametric or semi-parametric model for the angular measure.
Both methods involve nontrivial computation, although our method
requires only the numerical computation of a one-dimensional integral,
whereas Wang and Stoev's approach requires computation in fitting an
adequate discrete approximation to the spectral measure and then in
drawing from the conditional distribution.

\subsection{Approximating the conditional density when observations are
large via the angular measure}
\label{secpredViaAngMsr}

Let $\mathbbm Z_{-d} = (Z_1, \ldots, Z_{d-1})^T$, and define $\mathbbm z_{-d}$
analogously. Working with the $L_1$ norm, our goal is to approximate
the distribution of $[Z_d \mid\mathbbm Z_{-d} = \mathbbm z_{-d}]$ when $\| \mathbbm
z_{-d} \| > r_*$ and $r_*$ is large.
Let us assume that $H$ is absolutely continuous with respect to the
Lebesgue measure on $S_{d-1}$ and let $h$ denote the corresponding density.

To approximate the conditional density, we employ the conditional
p.d.f.
%
\begin{equation}
\label{eqndensApprox}
f_{Z_d \mid\mathbbm Z_{-d}}(z_d \mid\mathbbm z_{-d})
\approx
\frac
{\| \mathbbm z \|^{-(d+1)}h ( {\mathbbm z}/{\| \mathbbm z \|} )}
{\int_0^\infty\|\mathbbm z(t)\|^{-(d+1)} h
( {\mathbbm z(t)}/{\| \mathbbm z(t) \|
} ) \,dt},
\end{equation}
where $\mathbbm z = (z_1, \ldots, z_{d-1}, z_d)^T$ and $\mathbbm z(t) = (z_1,
\ldots, z_{d-1}, t)^T$.
This approximation arises from the point process representation for a
regularly varying random vector as we show
below.
Consequently, the approximate conditional distribution utilizes the
angular measure $H$, which characterizes the dependence in $\mathbbm Z$'s
components when $\| \mathbbm Z \|$ is large.

The first step in justifying the approximation (\ref{eqndensApprox})
is to characterize the limiting measure $\Lambda(\cdot)$ in terms of
Cartesian rather than polar coordinates.
%
\begin{proposition}
\label{thmmeasureCartesian}
Assume $\mathbbm Z$ is $d$-dimensional multivariate regularly varying with
common marginal distributions, tail index $\alpha= 1$, and angular
density $h$.
Let $N_n$ denote the sequence of point processes consisting of the
point masses located at $\{\mathbbm Z_i/a_n, i = 1, 2, \ldots, n\}$, where
$\mathbbm Z_i$ are i.i.d. copies\vadjust{\goodbreak} of $\mathbbm Z$, and let $N$ be the limiting point
process as $n \rightarrow\infty$. Denote the intensity measure of $N$
by $\Lambda(\cdot)$. Then $\Lambda(d\mathbbm z) = \| \mathbbm z \|^{-(d+1)} h(\mathbbm
z \| \mathbbm z \|^{-1}) \,d\mathbbm z$.
\end{proposition}
\begin{pf}
The proof is a simple change-of-variables argument.
Define the transformation $p\dvtx (0,\infty) \times S_{d-1} \mapsto\cal
{C}$ by $\mathbbm z:= p(r, \mathbbm w) = r \mathbbm w $ and note that $p$ is the
inverse of the usual Cartesian-to-polar coordinate transform.
To make the change of variables, we need $| {\det J_{p^{-1}}} |$.
It is known that $| {\det J_p} | = r^{(d-1)}$
[\citet{hogg2005}, Example 3.37 (specific for the $L_1$-norm) and \citet
{song1997lp}, Lemma 1.1 (for the general $L_q$-norm)].
Thus, $| {\det J_{p^{-1}}} | = \| \mathbbm z \|^{-(d-1)}$.

Let $A$ be an arbitrary set bounded away from $\{ \mathbbm0 \}$, and
consider $\Lambda(A)$:
\begin{eqnarray*}
\Lambda(A)
&=& \int_{(r, \mathbbm w) \in p(A)} r^{-2} h(\mathbbm w) \,dr\\
&=& \int_{\mathbbm z \in A} \| \mathbbm z \|^{-2} h(\mathbbm z \| \mathbbm z \|^{-1}) \|
\mathbbm z \|^{-(d-1)} \,d\mathbbm z\\
&=& \int_{\mathbbm z \in A} \| \mathbbm z \|^{-(d+1)} h(\mathbbm z \| \mathbbm z \|^{-1})
\,d\mathbbm z.
\end{eqnarray*}

Thus, $\Lambda(d\mathbbm z) = \| \mathbbm z \|^{-(d+1)} h(\mathbbm z \| \mathbbm z \|^{-1})
\,d\mathbbm z$.
\end{pf}
\begin{remark*}
The result is similar to Theorem 1 in \citet{coles91},
which allows one to start with a known multivariate max-stable
distribution with unit Fr\'echet marginals and find its corresponding
angular measure.
Coles and Tawn state ``the drawback to the use of theorem 1 is that it
can be applied only to MEVDs, of which very few have been obtained''
(page 381).
It is important to note that our aim is somewhat the reverse of Coles
and Tawn: we wish to start with an angular measure, and obtain an
approximation for the (conditional) density given the observed values
are large.

Now, for any $r_0 > 0$ and $\mathbbm z \in\reels^d$ such that $\| \mathbbm z \|
> r_0$, let
\[
F_{\mathbbm Z/ a_n}(\mathbbm z, r_0) =
\Proba
\biggl(
\frac{\mathbbm Z}{a_n} \in[\mathbbm z, \bolds\infty) \Bigm| \frac{\| \mathbbm Z \|}{a_n} > r_0
\biggr).
\]
Then,
%
\begin{eqnarray}
F_{\mathbbm Z/a_n}(\mathbbm z, r_0)
&=& \frac{
\Proba( ({\mathbbm Z}/{a_n}) \in[\mathbbm z, \bolds\infty),
({\| \mathbbm Z \|
}/{a_n}) > r_0 )}
{\Proba( ({\| \mathbbm Z \|}/{a_n}) > r_0 )
}\nonumber\\
&=& \frac{n \Proba( ({\mathbbm Z}/{a_n}) \in[\mathbbm z, \bolds\infty) )}
{n \Proba( ({\| \mathbbm Z \|}/{a_n}) > r_0 )}\nonumber\\
&\rightarrow& \frac{
\Lambda( [\mathbbm z, \bolds\infty) )}
{\Lambda( \{ \mathbbm z \mid\| \mathbbm z \| > r_0 \} )
} \qquad\mbox{from (\ref{eqnmvRegVarSeq})} \nonumber\\
\label{eqnstartingH}
&=& r_0 \Lambda( [\mathbbm z, \bolds\infty) ) \qquad\mbox{because }\int_{r > r_0}
r^{-2} \,dr = r_0^{-1}\\
\label{eqnFnz}
&=& r_0 \int_{[\mathbbm z, \bolds\infty)} \| \mathbbm z \|^{-(d+1)} h(\mathbbm z \| \mathbbm
z \|^{-1}) \,d\mathbbm z \qquad\mbox{see Proposition~\ref{thmmeasureCartesian}.}
\end{eqnarray}

We wish to speak of $f_{\mathbbm Z/a_n}(\mathbbm z, r_0)$, a joint density of
$\mathbbm Z/a_n$ given $\| \mathbbm Z \| /a_n > r_0$.
Heuristically from (\ref{eqnFnz}), we will assume that
%
\begin{equation}
\label{eqndensity}
f_{\mathbbm Z/a_n}(\mathbbm z, r_0) \rightarrow r_0 \| \mathbbm z \|^{-(d+1)} h(\mathbbm z
\| \mathbbm z \|^{-1})\qquad \mbox{for } \| \mathbbm z \| > r_0
\end{equation}
as $n \rightarrow\infty$.
More specifically, the convergence
would be guaranteed if $\frac{d}{d\mathbbm z} F_{\mathbbm
Z/a_n}(\mathbbm z$, $r_0)$ converged uniformly to $r_0 \| \mathbbm z
\|^{-(d+1)} h(\mathbbm z \| \mathbbm z \| ^{-1})$, allowing us to
switch the order of the limits associated with differentiation and as
$n \rightarrow\infty$.
See also Theorem 6.4 in \citet{Resnick07} in which regularly varying
densities are described.
\end{remark*}
\begin{example*}[(Bivariate logistic distribution)]
Let\vspace*{1pt} $\mathbbm Z$ have c.d.f. $\Proba(Z_1 \leq z_1, Z_2 \leq z_2) = \exp[ - (
z_1^{-1/\beta} + z_2^{-1/\beta} )^{\beta} ]$ for $\beta\in(0, 1]$.
$\mathbbm Z$ is then said to have a bivariate logistic distribution, which
is a known multivariate max-stable distribution, and which (more
importantly for our purposes) is also regularly varying with common
unit-Fr\'echet marginals $\Proba(Z_j \leq z) = \exp(-z^{-1})$ for $j =
1, 2$.
\citet{coles91} show that the angular density of the bivariate logistic
is given by
%
\[
h(\mathbbm w) = \frac{1}{2} \biggl( \frac{1}{\beta} - 1\biggr) (w_1 w_2)^{-1/\beta- 1}
( w_1^{-1/\beta} + w_2^{-1/\beta} )^{\beta- 2}.
\]
For a bivariate regularly varying random vector with unit Fr\'echet
margins, it can be shown that $a_n = 2n$ is a normalizing sequence such
that $\Proba(\| \mathbbm Z \| > a_n) \sim n^{-1}$.
Now,
%
\begin{eqnarray}\label{eqninclusionexclusion}
\Proba\biggl( \frac{\mathbbm Z}{2n} \in[ \mathbbm z, \bolds\infty)\biggr)
&=& \Proba( Z_1 > 2nz_1, Z_2 > 2nz_2) \nonumber\\
&=& 1 - \exp(- (2nz_1)^{-1}) - \exp(- (2nz_2)^{-1}) \nonumber\\
&&{} + \exp\bigl[ - \bigl( (2nz_1)^{-1/\beta} + (2nz_2)^{-1/\beta} \bigr)^{\beta} \bigr]
\\
&=& (2nz_1)^{-1} + (2nz_2)^{-1} -
\bigl(
(2nz_1)^{-1/\beta} + (2nz_2)^{-1/\beta}
\bigr)^{\beta}\nonumber\\
&&{} + o(n^{-1}) \nonumber
\end{eqnarray}
and, hence, for $\| \mathbbm z \| > r_0$,
\[
F_{\mathbbm Z/2n}(\mathbbm z, r_0)
\rightarrow
\tfrac{1}{2} r_0 \bigl( z_1^{-1} + z_2^{-1} - (z_1^{-1/\beta} + z_2^{-1/\beta
})^\beta\bigr).
\]
Differentiating this, we obtain the density
\begin{eqnarray*}
f_{\mathbbm Z/2n}(\mathbbm z, r_0)
&\rightarrow&
\frac{1}{2} r_0 ( \beta^{-1} - 1 )
( z_1^{-1/\beta} + z_2^{-1/\beta} ) ^{\beta- 2}
z_1^{-1/\beta- 1} z_2^{-1/\beta- 1} \\
&=& \frac{1}{2} r_0 ( \beta^{-1} - 1 )(z_1 + z_2)^{-3}
\biggl( \biggl( \frac{z_1}{z_1 + z_2} \biggr)^{-1/\beta} +
\biggl( \frac{z_2}{z_1 + z_2} \biggr)^{-1/\beta} \biggr) ^{\beta- 2} \\
&&{} \times\biggl( \frac{z_1}{z_1 + z_2} \biggr)^{-1/\beta- 1}
\biggl( \frac{z_2}{z_1 + z_2} \biggr)^{-1/\beta- 1} \\
&=& r_0 \| \mathbbm z \|^{-3} h(\mathbbm z \| \mathbbm z \|^{-1}),
\end{eqnarray*}
which agrees with (\ref{eqndensity}).
Similar arguments could be made for logistic models of dimension $d >
2$, but the inclusion/exclusion argument made in (\ref{eqninclusionexclusion}) becomes tedious.

Now, let us assume $n$ is fixed, but large enough such that $f_{\mathbbm Z/a_n}(\mathbbm z, r_0) \approx r_0 \| \mathbbm z \|^{-(d+1)} h(\mathbbm z \| \mathbbm z \|^{-1})$.
We wish to approximate $f_{\mathbbm Z}(\mathbbm z, r_*)$, the density of $\mathbbm Z$
given that $\| \mathbbm Z \| > r_*$ where $r_*$ is large.
To obtain an approximation, we do a change-of-variables from $\mathbbm Z/a_n$ to $\mathbbm Z$, which yields
%
\begin{eqnarray}
f_{\mathbbm Z}(\mathbbm z, r_*) &\approx& r_0 \| \mathbbm z/a_n
\|^{-(d+1)} h (\mathbbm z \| \mathbbm z \|^{-1})
a_n^{-d} \nonumber\\[-8pt]\\[-8pt]
&=& r_* \| \mathbbm z \|^{-(d+1)} h (\mathbbm z \| \mathbbm z
\|^{-1}),\nonumber
\end{eqnarray}
where $r_* = a_n r_0$, and thus is large.

Finally, consider the conditional distribution of $[Z_d \mid\mathbbm Z_{-d}
= \mathbbm z_{-d}]$ when $\| \mathbbm z_{-d} \| > r_*$ and $r_*$ is large.
Integrating to normalize the conditional density yields (\ref{eqndensApprox}).
\end{example*}
\subsection{An approximation example}
\label{secsimStudy}

\begin{figure}

\includegraphics{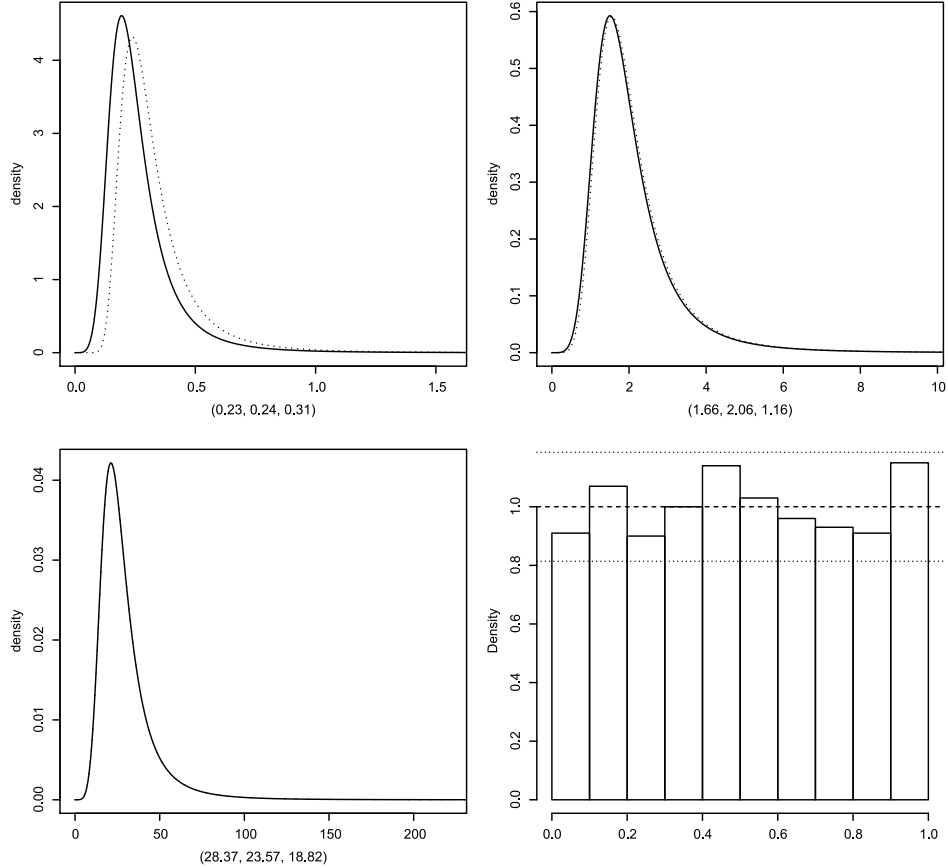}

\caption{Upper left, upper right and lower left panels show three
approximations of the conditional density of the third component of a
trivariate logistic random variable given the first two components. The
true conditional density is shown with the dotted line, the
approximated density with the solid line. Note the different scales for
the horizontal axis for these three figures. The approximation is poor
when the observed values are small (upper left), but improves as these
values become larger (upper right, lower left). The bottom right panel
shows the PIT histogram of the largest 1000 (as determined by $z_1 +
z_2$) of 5000 total simulations. As the PIT histogram is flat, it shows
that the approximation is good for these large observations. Dotted
lines indicate the approximate 0.05 and 0.95 quantiles for the sampling
distributions of each bin under the null hypothesis that the
conditional distribution is correct.}
\label{figlogisticSim}
\end{figure}

We investigate our approximation\break method via an example with a known
distribution and angular measure.
The trivariate logistic is a random vector with distribution
$
\Proba(Z_1 \leq z_1, Z_2 \leq z_2, Z_3 \leq z_3) = \exp[ -
(z_1^{-1/\beta} + z_2^{-1/\beta} + z_3^{-1/\beta})^\beta) ]
$
for $\beta\in(0, 1]$.
The angular measure of the trivariate logistic is given by
%
\begin{eqnarray}
\label{eqnangMsrLogistic} h(\mathbbm w) &=& \frac{1}{3}\biggl(
\frac{1}{\beta} - 1 \biggr) \biggl( \frac{2}{\beta} - 1 \biggr) ( w_1
w_2 w_3 )^{-1/\beta- 1}\nonumber\\[-8pt]\\[-8pt]
&&{}\times ( w_1^{-1/\beta} + w_2^{-1/\beta} +
w_3^{-1/\beta} )^{\beta-3}.\nonumber
\end{eqnarray}
We first investigate the quality of the approximation as $\| \mathbbm z_{-d}
\|$ increases and when $\beta= 0.3$. Since both the distribution and
the angular measure are known, we can compare the approximated
conditional density from (\ref{eqndensApprox}) to the actual
conditional density.
The first three panels of Figure~\ref{figlogisticSim}
show how the approximation improves as the magnitude of the
observations grows.
The top left panel shows that when the observed values are small $(z_1
= 0.23, z_2 = 0.24)$, the approximation to the true conditional density
is poor.
However, as the next two panels show, when the observations are
sufficiently large, the approximation is quite good.

Next we use a simulation experiment to assess the skill in using (\ref
{eqndensApprox}) for approximating the conditional density when the
conditioning observations are extreme.
From the \texttt{R} package \texttt{evd} [\citet{evd}], we simulate 5000
trivariate logistic random vectors with $\beta= 0.3$.
Let $\mathbbm Z_{i} = (Z_{i,1}, Z_{i,2}, Z_{i,3})^T$,\vspace*{1pt} $i = 1, \ldots,
5000$,
denote the i.i.d. random variables and $\mathbbm z_i = (z_{i,1}, z_{i,2},
z_{i,3})^T$ be the realized values, of which only $z_{i,1}$ and
$z_{i,2}$ are initially observed.
We rank the realizations $\mathbbm z_i$ according to the sum of the observed
values $z_{i,1} + z_{i,2}$.
We then apply our approximation method to the largest 1000 of these
simulations which corresponds to the condition $z_{i,1} + z_{i, 2} > 8.7$.
As each simulated random vector results in a unique conditional density
approximation, we assess our method via a probability integral
transform (PIT) or rank histogram [\citet{gneiting2007b}, \citet
{Wilks06}, Section 7.7.2].
Let $f_{Z_{i,3} \mid Z_{i,1}, Z_{i,2}}(z_{i,3} \mid z_{i,1}, z_{i,2})$
be the approximated conditional density given by (\ref{eqndensApprox}).
On simulation $i$, if the observed values are large enough, we let $p_i
= \int_{-\infty}^{z_{i,3}} f_{Z_{i,3} \mid Z_{i,1}, Z_{i,2}}(s \mid
z_{i,1}, z_{i,2}) \,ds$, where $z_{i,3}$ is the (previously unobserved)
value of $Z_{i,3}$.
We then construct a histogram for the values $p_i$.
If the approximation is well-calibrated, then the PIT histogram should
be flat, since there should be equal probability of $p_i$ occurring in
each bin.
If the conditional density were correct, the counts in each bin would
have a binomial $(n = 1000, p = 0.1)$ distribution and approximate
quantiles for the sampling distribution can be generated under this
null hypothesis.
The bottom right panel indicates that the approximation seems to be
quite good given that the observations are large, and given that the
angular measure is known.

\section{Application to nitrogen dioxide air pollution measurements}
\label{secapp}

The nitrogen oxides (NO$_x$) constitute one of the six common air
pollutants for which the US EPA is required to set air quality
standards by the Clean Air Act.
Of the various nitrogen oxides, nitrogen dioxide (NO$_2$) is the
component of ``greatest interest'' and is used as an indicator of the
entire group of NO$_x$.\footnote{\url{http://www.epa.gov/air/nitrogenoxides/}.}
According to the EPA fact sheet [\citet{NO2FactSheet}], short term
NO$_2$ exposures have been shown to cause adverse respiratory effects
such as increased asthma symptoms.
In January 2010, a new 1-hour NO$_2$ standard was set at 100 parts per
billion (ppb) to protect against adverse health effects due to
short-term exposure to NO$_2$.

Under the guidelines set by the EPA's Ambient Air Monitoring
Program,\footnote{\url{http://epa.gov/airquality/qa/monprog.html}.} state and
local agencies are charged with establishing and maintaining a network
of air pollution monitoring stations.
The EPA has made available data from these stations.
Using an online
tool,\footnote{\url{http://www.epa.gov/airdata/}.} we collected data from five stations located in Washington DC
and nearby Virginia which were all active during the entire period from
1995--2010.
The stations were Alexandria (site ID: 51-510-0009), McMillan
(11-001-0043), River Terrace (11-001-0041), Takoma School (11-001-0025)
and Arlington (51-013-0020).
The locations of these stations are shown in Figure~\ref{figmonitorLocs}.

Of course, air pollution measurements are of most interest when levels
are believed to be high, and monitors only record pollutant levels at
specific locations.
We test our prediction method when observations are large on these five
Washington-area stations.
We aim to predict the NO$_2$ level at the Arlington station, given the
NO$_2$ measurement at the other four stations.
We choose NO$_2$ because all five stations measured this pollutant, and
NO$_2$ appears to have the heaviest tail of the pollutants we examined.


From each of the five stations, we extract the daily maximum NO$_2$
measurement; all of the stations have over 5000 daily NO$_2$
measurements recorded between Jan 1, 1995--Jan 31, 2010 which meet
EPA's daily summary quality requirements.
From these, we keep only days in which all five stations have
measurements, resulting in 4497 daily measurements.
Finally, because the data are truncated to the nearest ppb,
the empirical c.d.f. appears quite discrete.
Thus, we add a uniform random variable on the interval $[-0.5, 0.5]$ to
the data so that they behave more like the underlying continuous
variable.\footnote{Data available at
\texttt{
\href{http://www.stat.colostate.edu/\textasciitilde cooleyd/DataAndCode/PredExtremes/}{http://www.stat.colostate.edu/\textasciitilde cooleyd/DataAndCode/}
\href{http://www.stat.colostate.edu/\textasciitilde cooleyd/DataAndCode/PredExtremes/}{PredExtremes/}}.}

\begin{figure}

\includegraphics{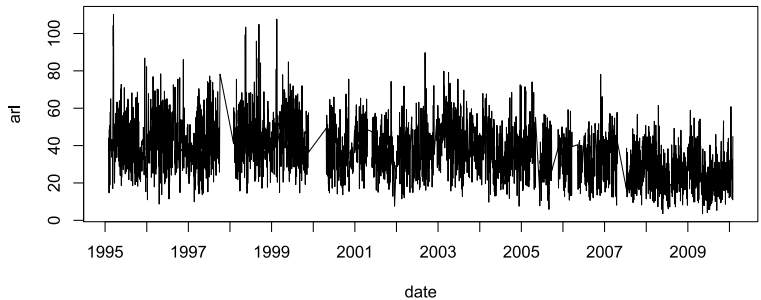}

\caption{Time series plot of the retained measurements at the
Arlington station.}
\label{figarlTS}
\end{figure}

\begin{figure}

\includegraphics{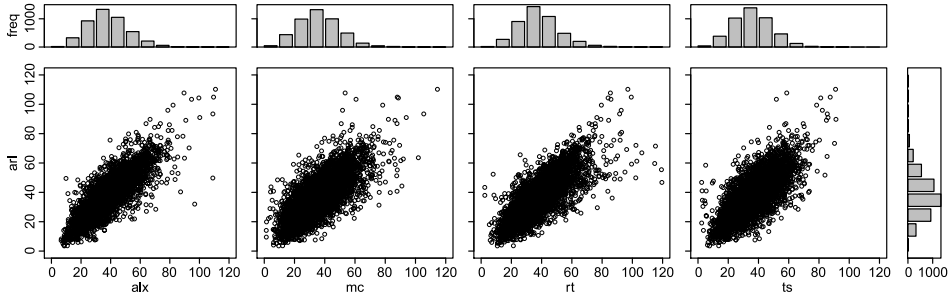}

\caption{Scatterplots of the measurements at the Arlington (arl)
station verses the other four stations.}
\label{figarlScatter}
\end{figure}

Figure~\ref{figarlTS} shows the time series of the retained
measurements at the Arlington station.
Unlike other pollutants such as ground-level ozone, there does not
appear to be a strong seasonal effect for NO$_2$.
Although a very weak seasonal signal is detectable for a moving-average
smoothed time series, this signal is hard to discern from a smoothed
periodogram.
%
It also appears that NO$_2$ levels have decreased at this site over the
study period, and the other stations show a similar, but weaker, trend.
Checking for serial dependence in the data, we find the sample
autocorrelation function of the deseasonalized data shows a highly
significant correlation only at lag 1 [$\hat\rho(1) = 0.35$].
Figure~\ref{figarlScatter} shows scatter plots of the measurements at
the Arlington station versus the four stations used for prediction.
The strong positive correlation between NO$_2$ measurements shown in
Figure~\ref{figarlScatter} is indicative of that found among all
pairs, with the strongest sample correlation being 0.83 between
Arlington and Alexandria and the weakest being 0.66 between Alexandria
and McMillan.
Figure~\ref{figarlScatter} also shows that largest values can be
coincident between stations.
In our analysis that follows, we assume that the dependence in the
upper tail of the joint distribution of NO$_2$ measurements is not
affected by the weak seasonality or the trend found in the data.
We checked the trend assumption by fitting angular measure models to
the first half and second half of the multivariate time series
separately and found similar parameter estimates.
We also ignore the serial dependence in the data, predicting the
Arlington station's measurement using only the other four stations'
measurements from that day.

Let $\mathbbm Y_t = (Y_{t,1}, \ldots, Y_{t,5})^T$ represent the random
vector of measurements on day $t$ at the five locations.
Our first task is to estimate the angular measure which describes the
tail dependence of the NO$_2$ measurements at these locations.
As formulated in Section~\ref{secmvRegVar}, angular measure models
assume a common marginal distribution with tail index $\alpha= 1$.
To obtain common marginal distributions, we use the following procedure.
Mean residual life plots [\citet{Coles01a}, Section 4.3.1] are used to
select the 0.93 quantile as an appropriate threshold above which each
location's data were approximately Pareto-distributed.
At each location, a generalized Pareto distribution (GPD) is fit to the
data above the threshold via maximum likelihood.
Letting $\hat F_j$ be the estimated marginal distribution function
formed by using the empirical distribution below the threshold and the
fitted GPD above (appropriately weighted by the observed exceedance
probability), each location's data are transformed to have a unit Fr\'
echet distribution: $Z_{t,j} = (-\log(\hat F_j(Y_{t,j}))^{-1}$.
Table~\ref{tabGPD} summarizes the marginal tail estimates.

\begin{table}
\tablewidth=240pt
\caption{Threshold and GPD estimates (and standard errors) of the tail
for the five Washington DC area locations. The GPD is parametrized $P(Y
> y | Y > u) = (1 + \xi\frac{y - u}{\psi})_+^{-1/\xi}$.
If~$\xi> 0$,~the~tail~index~$\alpha= 1/\xi$}
\label{tabGPD}
\begin{tabular*}{\tablewidth}{@{\extracolsep{\fill}}c c c c c@{}}
\hline
\textbf{Site \#} & \textbf{Location} & $\bolds{u}$
& $\bolds{\hat\psi}$ & $\bolds{\hat\xi}$ \\
\hline
1 & alx & 59.44 & 7.78 (0.64) & 0.07 (0.06)\\
2 & mc & 56.80 & 8.29 (0.70) & 0.05 (0.06)\\
3 & rt & 59.69 & 8.96 (0.78) & 0.10 (0.07)\\
4 & ts & 55.51 & 6.67 (0.55) & 0.02 (0.06)\\
5 & arl & 57.97 & 7.56 (0.62) & 0.07 (0.06)\\
\hline
\end{tabular*}
\end{table}

The data are divided into a training set and test set.
The training set, consisting of two-thirds of the available data
($n_{\mathrm{train}} = 2998$), is used to fit a five-dimensional angular measure model.
The test set, consisting of the other one-third ($n_{\mathrm{test}} = 1499$), is
used to test the prediction method.
Because of the decreasing trend, we construct the test set by
extracting every third observation so that both the training and test
sets would reflect the behavior over the entire study period.

The pairwise beta model [\citet{cooley10b}] is an angular measure model
for dimension $d > 2$ with parameters which help to control the amount
of dependence between each pair of elements in the random vector.
We fit the pairwise beta via maximum likelihood, and the likelihood
arises by assuming that the point process relationship implied by (\ref
{eqnmvRegVarPolar}) is exact for large observations [\citet{coles91},
\citet{cooley10b}].
The largest observations were determined by $\| \mathbbm z_t \|$, that is,
the radial component of the transformed data, and the largest 210
observations (0.93 quantile) were used to fit the model.

The pairwise beta has angular density given by
%
\begin{equation}
\label{eqnpbSpecMsr}
h(\mathbbm w; \gamma, \bolds\beta) =
K_d(\gamma) \sum_{1 \leq j < k \leq d} h_{j,k} (\mathbbm w; \gamma, \beta_{j,k})
\qquad\mbox{for } 0 < w_j < 1
\end{equation}
where
\begin{eqnarray*}
h_{j,k}(\mathbbm w; \gamma, \beta_{j,k}) &=&
(w_j + w_k)^{2\gamma-1}
\bigl(1 - (w_j + w_k)\bigr)^{\gamma(d-2)-d+2} \\
&&{} \times\frac{\Gamma(2 \beta_{j,k})}{\Gamma^2(\beta_{j,k})}
\biggl(\frac{w_j}{w_j + w_k}
\biggr)^{\beta_{j,k} - 1}\biggl(\frac{w_k}{w_j + w_k}\biggr)
^{\beta_{j,k} - 1}
\end{eqnarray*}
and
\[
K_d(\gamma) = \frac{2(d-3)!}{d(d-1)\sqrt{d}}
\frac{\Gamma(\gamma d + 1)}{\Gamma(2\gamma+ 1)\Gamma(\gamma(d-2))}
\]
is a normalizing constant.
The estimated parameters for the fitted model are given in Table \ref
{tabpbFit}.
In the pairwise beta model the magnitude of the $\beta_{i,j}$ parameter
is related to the level of dependence between the $i$th and $j$
components; the fact that $\hat\beta_{1,5}$ is the largest indicates
that the Alexandria and Arlington stations show the strongest tail dependence.

\begin{table}
\tabcolsep=5pt
\caption{Parameter estimates (and standard errors) for the pairwise
beta angular measure model fit to the Washington DC NO$_2$ data}
\label{tabpbFit}
\begin{tabular*}{\tablewidth}{@{\extracolsep{\fill}}c c c c c c c c c c c@{}}
\hline
$\bolds{\hat\gamma}$ & $\bolds{\hat\beta_{1,2}}$
& $\bolds{\hat\beta_{1,3}}$ & $\bolds{\hat\beta
_{1,4}}$ & $\bolds{\hat\beta_{1,5}}$
& $\bolds{\hat\beta_{2,3}}$ & $\bolds{\hat\beta_{2,4}}$
& $\bolds{\hat\beta_{2,5}}$ & $\bolds{\hat
\beta_{3,4}}$
& $\bolds{\hat\beta_{3,5}}$ & $\bolds{\hat\beta_{4,5}}$\\
\hline
0.37 & 0.51 & 0.64 & 0.56 & 6.11 & 0.76 & 1.64 & 0.96 & 0.56 & 0.98 &
1.01\\
(0.03) & (0.18) & (0.28) & (0.19) & (2.59) & (0.44) & (1.08) & (0.51)
& (0.20) & (0.51) & (0.61)\\
\hline
\end{tabular*}
\end{table}

\begin{figure}

\includegraphics{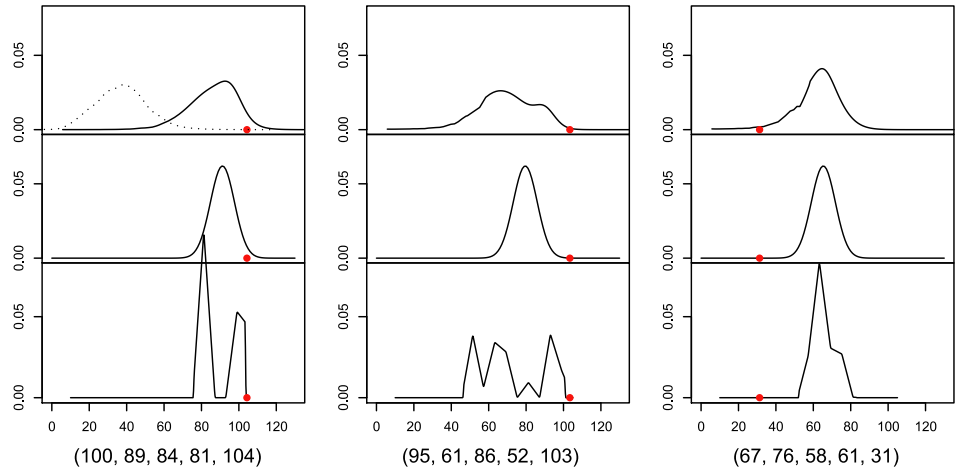}

\caption{Comparison of the approximated conditional densities at the
Arlington site given the measurements at the other sites: $g_{Y_{t,5} |
\mathbbm Y_{t,-5}} (y_{t,5} | \mathbbm y_{t,-5})$ for three different days with
high measurements. Top figure in each column is the approximated
conditional density via the angular measure, middle figure is from
simple kriging, and the bottom figure is from indicator kriging. Below
each figure is the vector of actual measurements at all five sites. The
fifth element corresponds to the Arlington site which we are trying to
predict and which is plotted with a dot in the figures. The dotted line
in the upper left corresponds to the marginal distribution for the
Arlington site.}
\label{figthreeExamples}\vspace*{-3pt}
\end{figure}

For the test set, we assume that the Arlington station is not observed,
and aim to approximate the conditional density of this station's NO$_2$
measurement given the measurements at the other four stations.
Since our method is only valid when the observations are large, we
perform prediction for the 105 test-set observations with the largest
values of $\| \mathbbm z_{t, -5} \|$.
That is, we threshold at the empirical 0.93 quantile of the radial
component (sum) of the transformed data at the observed locations.
Using the fitted pairwise beta angular measure, the conditional density
$f_{Z_{t,5} | \mathbbm Z_{t, -5}}(z_{t,5} \mid\mathbbm z_{t,-5})$ was
approximated using the procedure described in
Section~\ref{secpredViaAngMsr} for each of these top 105
observations.\footnote{The code used to produce the results is
available at \texttt{%
\href{http://www.stat.colostate.edu/\textasciitilde cooleyd/DataAndCode/PredExtremes/PredExtremesFiles.zip}%
    {http://www.stat.}
\href{http://www.stat.colostate.edu/\textasciitilde cooleyd/DataAndCode/PredExtremes/PredExtremesFiles.zip}%
    {colostate.edu/\textasciitilde cooleyd/DataAndCode/PredExtremes/PredExtremesFiles.zip}}.} The
integration in the denominator of (\ref{eqndensApprox}) was
approximated using Simpson's Rule. These were then back-transformed to
obtain the conditional densities on the original scale $g_{Y_{t,5} |
\mathbbm Y_{t, -5}}(y_{t,5} \mid\mathbbm y_{t,-5})$. Three of the
approximated conditional densities can be found in the top row of
Figure~\ref{figthreeExamples}.

We compare our prediction method to two other approaches: best linear
unbiased prediction (kriging) and indicator kriging [\citet{Cressie93},
\citet{Schabenberger05}].
Kriging is a prediction method that utilizes only mean and covariance
information.
At its most fundamental level, kriging does not make a distributional
assumption, it provides a point prediction which corresponds to the
best linear unbiased predictor in mean-square prediction error (MSPE),
and additionally provides an estimate of the MSPE.
To obtain confidence intervals, typically a Gaussian assumption is made.
Furthermore, if one assumes the data arise from a Gaussian process,
then the kriging estimate and MSPE correspond to the conditional
expectation and variance.
Since our method generates a conditional distribution, we will compare
it to the conditional distribution provided by kriging under a Gaussian
assumption.

Our kriging procedure is in parallel to the angular measure procedure above.
The training set is used to\vadjust{\goodbreak} formulate a model; here all 2298
observations are used to estimate the mean NO$_2$ levels at all five
locations as well as to estimate the covariance matrix between the
measurements at the locations.
It is important to note that no spatial covariance function is fit, as
our training set allows us to estimate the covariance matrix directly.
Treating the mean and covariance as known, we then use simple kriging
[\citet{Cressie93}] to obtain a point prediction at the Arlington location
given the measurements at the other locations for the same 105 large
observations in the test set.
The MSPE is calculated from the estimated covariances, we use it to
obtain the estimated conditional density at the Arlington location
given the other measurements under a Gaussian assumption.

We also compare to indicator kriging [\citet{Cressie93},
\citet{Schabenberger05}] which is a nonparametric version of kriging designed
to provide estimates of $\Proba(Y_d > u | Y_1, \ldots, Y_{d-1})$ for a
given threshold $u$. When performing indicator kriging, one first needs
an estimate of the covariance matrix of the random variables
corresponding to the indicators $\mathbb{I}(Y_j > u)$ for all the $j$
locations. At time $t$, given observations $y_{t,1}, \ldots,
y_{t,d-1}$, these are converted into indicators, $\mathbb{I}(y_{t,j} >
u)$ and ordinary kriging is used to estimate $E[\mathbb{I}(Y_{t,d} >
u) \mid\mathbb{I}(y_{t,1} > u), \ldots, \mathbb{I}(y_{t,d-1} > u)]
= \Proba(Y_{t,d} > u \mid\mathbb{I}(y_{t,1} > u), \ldots,
\mathbb{I}(y_{t,d-1} > u))$. Repeating the analysis for various values
of $u$ allows one to estimate a conditional distribution, although
there is no guarantee that the estimate will be monotonic.\vadjust{\goodbreak}

Our indicator kriging analysis is again parallel to the angular measure
and simple kriging analyses.
We let $u$ vary from 10--105 ppm with a step size of 0.25~ppm which
covers the range of observations.
The training set is used to estimate the covariance matrix of
indicators at the various levels of $u$, and then indicator kriging is
performed on each of the sets of observations in the test set.
To guarantee that the conditional distribution is monotonic, we then
perform a monotone quadratic smoothing spline regression [\citet
{meyer2008}] on the estimates $\Proba(Y_{t,d} > u \mid\mathbb
{I}(y_{t,1} > u), \ldots, \mathbb{I}(y_{t,d-1} > u))$ for all the
values of $u$.
Densities are obtained by differentiating the smoothing spline.

Conditional densities obtained by the angular measure method, simple
kriging and indicator kriging are shown in Figure
\ref{figthreeExamples} for three different days' data.
In these three figures the conditional density approximated via the
angular measure is less concentrated than the conditional density from
simple kriging, and that proves to be the case in general.
The angular measure can also be somewhat skewed or slightly bimodal
depending on the combination of the observed measurements.
Although indicator kriging is performed for each pollution level $u$,
the conditional density as approximated by indicator kriging is very
rough, as there are only four locations.

%
\begin{figure}[b]

\includegraphics{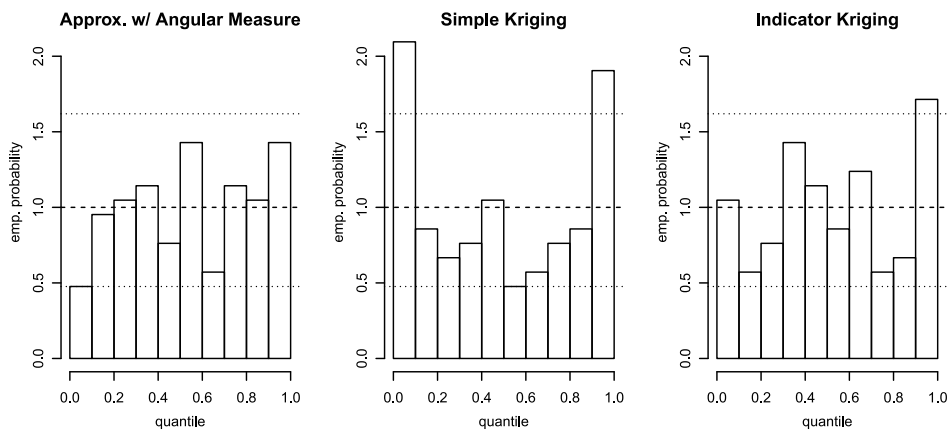}

\caption{PIT histograms for the angular measure approach (left),
simple kriging (center) and indicator kriging (right). Perfect
estimation of the conditional density would be indicated by a flat
histogram. Error bars are obtained for each decile from a binomial
distribution $(n = 105, p = 0.10)$.}
\label{figrankHistogram}
\end{figure}

We evaluate the performance of the three approaches using various methods.
All comparisons are done at the original scale.
To test the overall fit of the approximated conditional density, we
again use the PIT histogram.
Figure~\ref{figrankHistogram} shows the PIT histograms for all three methods.
The PIT histogram for the angular measure method is relatively flat
with perhaps some indication that the model is overestimating the
probability in the lower tail, resulting in too few observations\vadjust{\goodbreak}
falling in the first decile of the approximated conditional density.
This could be due to the angular density model: the pairwise beta model
fit to the data is certainly not the true model for the angular density
which is unknown.
It could also be due to threshold choice, although the parameter
estimates of the pairwise beta model did not appear to be sensitive to
the threshold.
The kriging estimate with the Gaussian assumption shows a classic
u-shape associated with underdispersion [\citet{Wilks06}, Section 7.7.2].
This model underestimates the probability of the observation occurring
in the lower tail and also the upper tail.
Indicator kriging also appears to underestimate the upper tail of the
distribution, resulting in
too many observations appearing in the highest decile of the
approximated conditional distribution.
Using the terminology of \citet{gneiting2007b}, the PIT histograms
indicate that the angular measure method is better (probabilistically)
calibrated, particularly in the upper quantiles of the predictive distribution.

Another performance evaluation is to see how well each method estimates
a quantile, and, particularly, a high quantile.
For instance, regulators might wish to have an accurate assessment of a
high quantile of an unmonitored location given large observations nearby.
Such an estimate could be used as a probabilistic upper bound, that is,
officials could state that they were 95\% confident that the level at
the unmonitored location was below a reported level.
For each of the 105 large observations, we use the approximated
conditional density from all three methods to estimate the 0.99, 0.95,
0.90, 0.75 and 0.50 quantiles.
We examine coverage by calculating the proportion of actual
observations that fell beneath these quantiles and also calculate each
method's quantile verification score (QVS) [\citet{gneiting2007}].
Let $g^{(m)}_{Y_{t,5} \mid\mathbbm Y_{t, -5}}(s \mid\mathbbm y_{t, -5})$ and
$G^{(m)}_{Y_{t,5} \mid\mathbbm Y_{t, -5}}(s \mid\mathbbm y_{t, -5})$ be
the\vspace*{1pt}
predictive density and cumulative distribution function for $Y_{t,5}
\mid\mathbbm Y_{t, -5} = \mathbbm y_{t, -5}$, where $m$ denotes the angular
measure method $(m = 1)$, kriging $(m = 2)$ or indicator kriging $(m = 3)$.
We have parameterized the QVS as in \citet{friederichs07},
\[
QVS^{(m)} = \sum_{t = 1}^{105} \rho_\tau\bigl(y_{t,5} - q_{t,\tau}^{(m)}\bigr),
\]
where
$q_{t,\tau}^{(m)} = G^{(m)\leftarrow}_{Y_{t,5} \mid\mathbbm Y_{t, -5}}(\tau)$,
and $\rho_\tau(u) = \tau u I(u \geq0) + (\tau- 1) u I(u < 0)$.
A~lower QVS score indicates better skill, but the scale of the QVS
score depends on the quantile to which it is being applied.
The QVS is a proper scoring rule, meaning that it is minimized if the
predictive distribution corresponds to the ``true'' distribution.
Both the coverage and QVS results for the tested quantiles are shown in
Table~\ref{tabhighQuantileEst}, as well as the sampling error
assuming independent Bernoulli trials with $p$ equal to the given quantile.
The angular measure method does a superior job of estimating the high
quantiles (0.99, 0.95 and 0.90) when the observations are large,\vadjust{\goodbreak}
whereas both simple kriging and indicator kriging underestimate these
high quantiles.
The angular measure method seems to be outperformed by indicator
kriging for the 0.75 and 0.50 quantiles, although its coverage rates
fall well within acceptable ranges when sampling error is accounted for.



\begin{table}
\tabcolsep=0pt
\caption{Gives the skill of the different methods for assessing high
quantiles. Coverage (Cvg) column reports the proportion of the
observations at the Arlington location that fell beneath the quantile
as calculated from the estimated conditional density and QVS column
reports the quantile verification~score (lower is better)}
\label{tabhighQuantileEst}
\begin{tabular*}{\tablewidth}{@{\extracolsep{\fill}}
l c c c c c c c c c c@{}}
\hline
& \multicolumn{10}{c@{}}{\textbf{Quantile}} \\[-4pt]
& \multicolumn{10}{c@{}}{\hrulefill}\\
& \multicolumn{2}{c}{\textbf{0.99}} & \multicolumn{2}{c}{\textbf{0.95}} &
\multicolumn{2}{c}{\textbf{0.90}} & \multicolumn{2}{c}{\textbf{0.75}}
& \multicolumn{2}{c}{\textbf{0.50}}\\[-4pt]
& \multicolumn{2}{c}{\hrulefill} & \multicolumn{2}{c}{\hrulefill}
& \multicolumn{2}{c}{\hrulefill} & \multicolumn{2}{c}{\hrulefill}
& \multicolumn{2}{c@{}}{\hrulefill}\\
& \textbf{Cvg} & \textbf{QVS} & \textbf{Cvg} & \textbf{QVS}
& \textbf{Cvg} & \textbf{QVS} & \textbf{Cvg} & \textbf{QVS}
& \textbf{Cvg} & \textbf{QVS}\\ \hline
Angular  & 0.97 & 40.97 & 0.93 & 134.77 & 0.88 & 225.68 & 0.70
& 398.97 & 0.44 & 502.51\\
\quad measure\\
Simple  & 0.92 & 65.80 & 0.83 & 170.04 & 0.81 & 246.26 & 0.65 &
378.27 & 0.54 & 444.84\\
\quad kriging\\
Indicator  & 0.90 & 67.80 & 0.86 & 153.41 & 0.83 & 238.63 &
0.73 & 377.20 & 0.49 & 452.68\\
\quad kriging\\
Sampling  & (0.01) & -- & (0.02) & -- & (0.03) & -- & (0.04) & --
& (0.05)\\
\quad error\\
\hline
\end{tabular*}
\end{table}

Each method's conditional density is essentially a probabilistic
forecast, and scoring rules have been developed which provide an
overall measure of the quality of probabilistic forecasts [\citet{gneiting2007}].
We assess the methods using two different proper scoring rules: the
logarithmic score and the continuous rank probability score (CRPS).
The logarithmic score for the prediction at time $t$ is given by $-\log
( g^{(m)}_{Y_{t,5} \mid\mathbbm Y_{t, -5}}(y_{t,5} \mid\mathbbm y_{t, -5}))$,
where $y_{t,5}$ is the actual observation at the Arlington station and
$g^{(m)}_{Y_{t,5}|\mathbbm Y_{t,-5}}$ is the estimated conditional density
via the angular measure approach $(m = 1)$, kriging $(m = 2)$ and
indicator kriging $(m = 3)$.
The logarithmic score has an information-theoretic basis and
corresponds to the Kullback--Leibler divergence between the predictive
density $g^{(m)}_{Y_{t,5} \mid\mathbbm Y_{t, -5}}(s \mid\mathbbm y_{t, -5}))$
and the Kronecker delta function $\delta_{s, y_{t,5}}$.
We assess the methods by the mean of the logarithmic scores
\[
\frac{1}{105}\sum_{t = 1}^{105} - \log\bigl( g^{(m)}_{Y_{t,5} \mid\mathbbm Y_{t,
-5}}(y_{t,5} \mid\mathbbm y_{t, -5})\bigr)
\]
over assumed independent realizations $\mathbbm y_t$, $t = 1, \ldots, 105$.
The mean logarithmic score is 3.93 for the angular measure approach,
4.24 for kriging, and infinity for indicator kriging, as 8 of the 105
of the observations $y_{t,5}$ fall outside the support of the\vadjust{\goodbreak}
predictive distribution.
Since a lower score is better, the angular measure method outperforms
kriging and indicator kriging by this performance measure.

The logarithmic score has been criticized, as it is not ``robust'' to
cases when observations fall outside the support of the distribution as
in the indicator kriging case above.
A popular alternative is the CRPS.
For our example, the CRPS for a particular day $t$ is given by
%
\begin{equation}
\label{eqncrps}
\int_{-\infty}^{\infty}
\bigl(
G^{(m)}_{Y_{t,5} \mid\mathbbm Y_{t,-5}}(s \mid\mathbbm y_{t, -5}) - \mathbb
{I}\{s \geq y_{t,5}\}
\bigr)^2 \,ds
\end{equation}
and can be understood as a nonlinear function of the area between each
method's predictive c.d.f. $G^{(m)}_{Y_{t,5} \mid\mathbbm Y_{t, -5}}(s \mid
\mathbbm y_{t, -5})$ 
and the heavyside function associated with the realized value $y_{t,5}$.
The CRPS score rewards appropriate centering of the predictive
distribution and narrowness of the predictive distribution otherwise
known as ``sharpness.''
We assess the three methods by the mean of the CRPS scores for $t = 1,
\ldots, 105$.
Given the PIT histograms and logarithmic scores, it is perhaps
surprising that the mean CRPS scores associated with the angular
measure method, kriging and indicator kriging are 6.83, 6.36 and 6.21,
respectively, indicating that by this performance measure, the angular
measure method is performing worst.
While all three methods produce predictive densities that are centered
(i.e., the realized values exceed the predictive density's median about
half of the time), the predictive densities from kriging and indicatior
kriging are sharper than (but not as well calibrated as) the density
produced by the angular measure method (Figure~\ref{figthreeExamples}).

\begin{figure}

\includegraphics{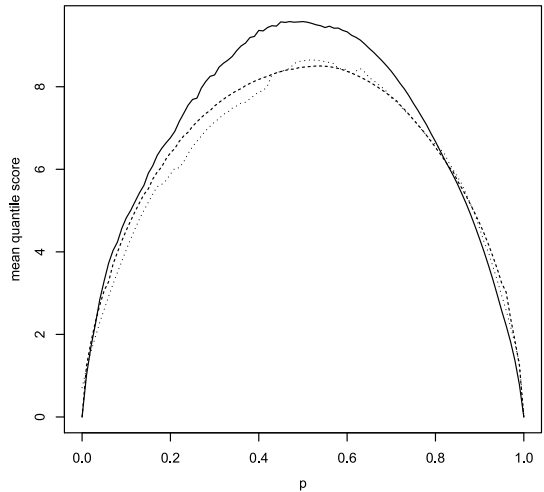}

\caption{Mean CRPS score for the three methods decomposed by quantile
$p$ as in Gneiting and Ranjan (\protect\citeyear{gneiting2011}). Solid line is the angular measure
method, dashed is kriging, dotted is ordinary kriging. The angular
measure method performs best for high quantiles, but performs less well
for the middle quantiles.}
\label{figquantileCRPS}\vspace*{-3pt}
\end{figure}

The CRPS score can be written as an integral with respect to a
threshold $s$ as in (\ref{eqncrps}) or, equivalently,\vspace*{1pt} in
terms of the quantile function ${G^{(m)}_{Y_{t,5} \mid\mathbbm
Y_{t,-5}} }{}^{-1}(p)$ and integrated with respect to $p \in(0,1)$
[\citet{gneiting2011}]. Further, the overall mean CRPS score can
be decomposed into a mean CRPS score at each $p$, then integrated with
respect to $p$. \citet{gneiting2011} suggest plotting the quantile
score verses $p$ as a diagnostic tool. When done for the three
forecasts, Figure~\ref{figquantileCRPS} shows that the angular measure
%
method outperforms the other methods for high quantiles, but both
kriging and indicator kriging outperform the angular measure method for
quantiles near 0.5, likely due to the increased sharpness of these
methods. Since the quantile scores are naturally larger near values of
0.5, the overall mean CRPS scores for kriging and indicator kriging end
up lower. \citet{gneiting2011} also discuss a quantile weighted
CRPS score
\[
\int_0^1 2 \bigl( \mathbb{I}\bigl\{y_{t,5} \leq{G^{(m)}_{Y_{t,5}
\mid\mathbbm Y_{t,-5}} }{}^{-1}(p) \bigr\} - p \bigr) \bigl({G^{(m)}_{Y_{t,5}
\mid\mathbbm Y_{t,-5}} }{}^{-1}(p) - y_{t,5} \bigr) v(p) \,dp,
\]
where one can choose the weight function $v(q)$ to emphasize quantiles
of interest.
Letting $v(q) = \mathbb{I} \{v(q) > 0.85 \}$, the mean weighted CRPS
scores for the angular measure method, kriging and indicator kriging
are 0.50, 0.57 and 0.55.\vadjust{\goodbreak}

\section{Summary and discussion}

In this work we obtain
an approximation for the distribution of a component of a regularly
varying random vector given that the observed components are large.
We apply the approximation to estimate the conditional distribution of
an air pollutant given nearby measurements that are large.
Results show that our method outperforms traditional spatial prediction
methods at capturing the conditional distribution of the random
variable when the observations are large.
PIT histograms show that our method is better calibrated, and the
method proves to be much better suited for obtaining probabilistic
upper bounds of the pollutant level.
For example, the estimated 95\% quantiles provided by kriging and
indicator kriging were too low and the actual exceedance rates were 17\%
and 14\%, respectively.
The exceedance rate of the angular measure method's estimated 95\%
quantile was 7\% and was within sampling error of 5\%.

We believe that this is the first work to perform prediction using
extremes techniques in a threshold exceedance setting.
The classic theory that leads to max-stable distributions and processes
is quite elegant and forms the foundation for all of extreme value theory.
Statistical practice utilizing multivariate max-stability generally
requires one to obtain component-wise block maximum data, and such data
can be viewed as ``artificial'' in the sense that one models data
vectors that are likely to have never occurred, since the block maxima\vadjust{\goodbreak}
are likely to occur at different times.
It seems natural to try and attempt to describe large concurrent
observations, and the framework of multivariate regular variation
allows this.

Our method relies on an adequate angular measure model. There has been
some renewed interest of late in constructing flexible models which
meet the moment conditions (\ref{eqncomCondition})
[\citet{cooley10b}, \citet{Ballani11}, \citet{Boldi07}].
However, no model with a finite parameterization can completely
describe the possible angular measures, and the existing models may not
prove to adequately model every multivariate data set. These models
become unwieldy as the dimension increases beyond moderate levels ($d
\approx5$). Certainly there remains a need for flexible multivariate
extremes models.
%

Although we apply our method to multivariate time series data, we do
not make use of any temporal dependence in the data.
Our method proceeds as if the sequence of multivariate random vectors
are i.i.d.
One could extend the method by allowing the marginal distributions to
vary in time; such extreme value models are regularly used [e.g.,
\citet{Beirlant04}, chapter~7] and might be required if, for instance, the
seasonality of this data had been more influential.

\printaddresses

\end{document}